# Incorporating Domain Knowledge into Health Recommender Systems using Hyperbolic Embeddings

Joel Peito, Qiwei Han[0000-0002-6044-4530]

Nova School of Business and Economics, Universidade NOVA de Lisboa, Campus de Carcavelos, 2775-405 Carcavelos, Portugal
`joel.depeito@gmail.com, qiwei.han@novasbe.pt`

**Abstract.** In contrast to many other domains, recommender systems in health services may benefit particularly from the incorporation of health domain knowledge, as it helps to provide meaningful and personalised recommendations catering to the individual's health needs. With recent advances in representation learning enabling the hierarchical embedding of health knowledge into the hyperbolic Poincaré space, this work proposes a content-based recommender system for patient-doctor matchmaking in primary care based on patients' health profiles, enriched by pre-trained Poincaré embeddings of the ICD-9 codes through transfer learning. The proposed model outperforms its conventional counterpart in terms of recommendation accuracy and has several important business implications for improving the patient-doctor relationship.

**Keywords:** Health Recommender Systems, Primary Care, Poincaré Embeddings, International Classification of Diseases, Patient-Doctor Relationship

## 1 Introduction

With the emergence of healthcare analytics and growing needs to leverage the prevalent electronic health records from healthcare providers, machine learning (ML) solutions, such as recommender systems (RS), have experienced growing relevance in the healthcare sector[1]. In fact, patients increasingly seek bespoke and digital medical solutions, similar to what they are used to from e-commerce and other domains. However, as patients' relationship to their doctors can be very personal and health conditions are sensitive topic, healthcare recommender systems (HRS) are subject to a different set of rules and evaluation criteria than other commercial applications of RS. For instance, product or movie RS do not operate under the same scrutiny regarding the reliability and trustworthiness of their predictions, since the ramifications of specific treatment or doctor recommendations are severer in nature.

In general, RS often capitalise on the target user's interaction data without the need of any additional information about the user itself or the recommended entity. While such methods can be highly performant, they usually do not offer a straightforward explanation as to *why* a specific product or movie is being recommended. Still, as long as users receive interesting recommendations, one can assume that this is not a particular issue for the latter. Patients, on the other hand, may be highly interested in solutions that not only fit their personal medical profile insofar, as they are built on medically meaningful information about the patient, but also provide explanations of the recommendation itself. That is to say, patients will arguably prefer recommendations optimised towards their individual medical needs, instead of recommendations based on the similarity to other patients that may show very similar behavioural patterns but have an entirely different medical background. Analogously, healthcare providers can treat this property as a value proposition to their clients, offering medically personalised recommendations and thereby meeting current market trends.

As such, this paper aims to investigate the possibility of adding such a medical personalisation dimension to the HRS by incorporating complex, domain-specific knowledge into the underlying model. More specifically, we propose a content-based RS for patient-doctor matchmaking built on real data from a leading European private healthcare provider. Patients' historical health records, as indicated by the ICD-9 codes[1] serve as the main source of domain knowledge. However, the use of ICD-9 code for encoding patients' health conditions faces a series of practical implementation problems. Chief among those is the structure of the data itself, as in nature ICD codes are encoded as hierarchical, tree-like structures that are hard to be embedded into the continuous space necessary for most ML models. Nevertheless, recent works [2,3] proposing hyperbolic embeddings for learning hierarchical representations appear to provide a bypass for this issue.

Consequently, we investigate how to incorporate complex domain knowledge, such as the ICD-9 hierarchy into a HRS using hyperbolic embeddings and examine whether such domain knowledge can add value to the HRS in terms of improving recommendation accuracy. For that purpose, we pursue the following approach: contextualising the topic, section 2 begins with a bibliographical examination of related works on HRS and lays out the benefits of embedding hierarchical data into the hyperbolic space. Notably, it will be shown why hyperbolic embeddings are inherently better equipped than their Euclidean counterparts to embed hierarchical, tree-like data into the continuous space. Moving forward, section 3 sheds light on the data at hand. Section 4 discusses the methods employed introducing the notion of hyperbolic distance as a similarity measure for RS and formulating two content-based models using said hyperbolic distance. For evaluation purposes, a conventional model is formulated to serve as a benchmark. Section 5 analyses the results of this investigation. Finally, section 6 discusses the conclusions we draw from this work, as well as suggestions for future research.

---

[1] International Classification of Diseases (ICD) is a comprehensive standard of diseases or medical conditions maintained by the WHO and widely used among healthcare organizations worldwide. It is revised periodically and now in its 10th version (known as ICD-10). However, the ICD-code used this study is still in the 9th version (ICD-9).

## 2 Background and related work

### 2.1 Recommender systems in healthcare

In general, RS are a subclass of information filtering systems with the goal to provide meaningful suggestions to users for certain items or entities, by attempting to predict the affinity or preference of a given user for said items [4]. RS can be broadly divided into three major categories: collaborative filtering (CF) approaches, content-based (CB) recommenders, and hybrid models, which are a combination of the former two. CF approaches rely solely upon past interactions recorded between users and items, whereas CB approaches use additional information about users and/or items [5]. More precisely, CF capitalises on behavioural data, *i.e.* users' co-occurrence patterns, in order to detect similar users and/or items and make predictions based on these similarities, while CB recommenders explore user or item metadata to derive user preferences and model the observed user-item interactions. Although CB recommenders do not suffer from the cold-start problem, i.e. the question of what to do with new users that have no prior interactions usable for predictions [6], CF approaches tend to outperform the former, as usually even a few ratings are more valuable than metadata about users or items [7]. Ultimately, a method to balance both CF and CB's respective limitations is to use hybrid recommenders, which are a combination of the former and the latter.

While RS have been widely used in e-commerce, e.g. for movie or product recommendations, within the healthcare domain RS are only recently emerging, due to elevated requirements regarding reliability and trustworthiness, as well as increased data privacy regulations [8,9]. The last years, however, have shown an increase in studies and research papers on HRS. Among those works, medical user profiling and medical personalisation have been particularly trending topics [10]. Hence, noteworthy examples of HRS applications include recommenders for relevant medical home care products [11], lifestyle adaption recommendations for hypertension treatment and prevention [12], identification of key opinion leaders [13], as well as clinical decision support systems using inherent methods of RS to capitalise on the large volume of clinical data [14]. Ultimately, [15] address the topic of patient-doctor matchmaking proposing RS for suggesting primary care doctors to patients based on their prior consultation history and metadata.

### 2.2 Hyperbolic embeddings

As has been hypothesized in the introduction, HRS might profit more than other areas from incorporating domain knowledge into the model. Within the healthcare context, such knowledge may include, for instance, a catalogue and categorisation of health conditions, such as the ICD-9 hierarchy. Abstracting a hierarchy into mathematical terms, it is essentially a complex tree that is defined as a connected graph in which for any pair of two vertices $u \neq v$ there is exactly one path connecting them [16]. An inherent characteristic of hierarchies or trees, however, is that they are discrete structures and thus embedding them in a way that can be used in machine learning models can be challenging, as the latter often rely on continuous representations [17]. Hence, the underlying question is, how to efficiently and accurately model an increasingly complex hierarchy - and accordingly an increasingly complex tree - into a continuous space, such that the information of the hierarchy can be used for machine learning?

Recent proposals by [2] and [3] suggesting hyperbolic embeddings to address this issue, have found much notice in the machine learning community. The rationale is that embeddings of hierarchical, tree-like data into the hyperbolic space perform better at the task of capturing and preserving the distances and complex relationships within a given hierarchy, than embeddings in the Euclidean space would. As a matter of fact, these works show that hyperbolic embeddings, even in very low dimensions, consistently outperform their higher-dimensional, Euclidean counterparts when learning hierarchical representations.

The reasons for said superiority lie within the properties of hyperbolic geometry itself. Hyperbolic space is a space with a constant negative curvature that expands exponentially rendering it inherently well-suited for the task of embedding a tree into the continuous space [2]. Meanwhile, the preferred geometrical models for representation learning tasks, such as the one at hand, are the Poincaré models as they offer to conform mapping between hyperbolic and Euclidean space, since angles are preserved – a convenient property when translating between spaces and models [17].

Recalling that the goal when embedding tree-like graphs into a continuous space is to preserve original graph distances, one needs to consider the hyperbolic distance:

$$d_H(x,y) = \text{acosh}\left(1 + 2\frac{||x-y||^2}{(1-||x||^2)(1-||y||^2)}\right) \tag{1}$$

In hyperbolic space, the shortest paths between two points, called geodesics, are curved (similarly to the space itself). Due to this curvature, the distance from the origin to a given point $d_H(O,x)$ grows towards infinity as $x$ approaches the edge of the disc, as can be observed in figure 1. Now, considering the embedding of a graph (or tree) into a continuous space, suppose $x$ and $y$ are children of a parent $z$, which is placed at the origin $O$. Then, the distance between $x$ and $y$ is:

$$d(x,y) = d(x,O) + d(O,y) \tag{2}$$

Normalizing this equation, provides the distance ratio of the original graph, i.e. $\frac{d(x,y)}{d(x,O)+d(O,y)} = 1$. This equation will be relevant in the following, since when comparing its behaviour in hyperbolic and Euclidean space, quite different effects can be observed. As is visualised in figure 1, when moving towards the edge of the unit disk, i.e. $x \rightarrow 1$, in Euclidean space $\frac{d_E(x,y)}{d_E(x,O)+d_E(O,y)}$ remains a constant, whereas in hyperbolic space $\frac{d_H(x,y)}{d_H(x,O)+d_H(O,y)}$ approximates 1, which is exactly the original graph distance ratio! Therefore, it can be seen that Poincaré embeddings are inherently better suited for this kind of representation learning task, due to their better capacity to preserve original graph distances with arbitrarily low distortion [17].

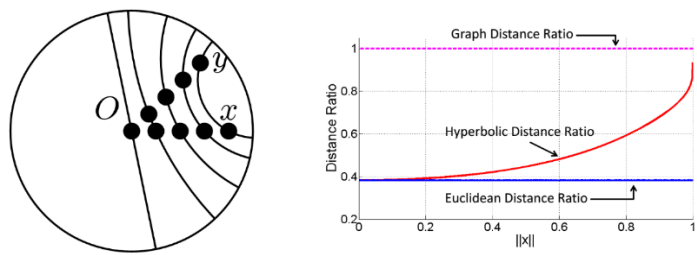

**Fig. 1.** The Poincaré disk model (left) and distance ratios of hyperbolic and Euclidean distance in comparison with original input graph distance ratio (right) [17].

While further analysis of the detailed mathematics of Poincaré embeddings, as laid out in [2], [3] and [17], are beyond the scope of this paper, we instead consider an actual use-case of Poincaré embeddings relevant to this work. For instance, [17] perform representation learning tasks for a variety of datasets, most of which related to NLP. In light of the given topic, however, their work on embeddings of the UMLS diagnostic hierarchy from ICD-9 vocabularies is of particular interest as they provide the very domain-specific knowledge needed for the proposed model.

## 3 Data

The dataset used in this work was provided by a leading European private heath network operating 18 hospitals or clinic centres across the country. Typically, data can be divided into three categories: 1) patients' demographic information, such as gender, age and home locations, and this information is further enriched with health records in the form of the ICD-9 code for inpatients, i.e. patients who stay at the hospital while under treatment; and 2) doctors' demographic and professional information, such as gender, age and the hospital they are working at and 3) the interactions between patients and doctors according to their consultation history.

Instead of learning the representation of ICD-9 code from our data, we resort to a transfer learning approach by relying on pre-trained Poincaré embeddings provided by [17]. In particular, they used the diagnostic hierarchy of ICD-9 vocabulary in the Unified Medical Language System Metathesaurus (UMLS) to retrieve Poincaré embeddings of medical concepts within the ICD-9 hierarchy. This method results in unique hyperbolic embeddings of medical concepts (identified by the CUI, *i.e.* Concept Unique Identifier) available in different levels of dimensionality (10, 20, 50 or 100d). We choose the 100d embeddings for our model, as [2] indicates that while Poincaré embeddings already perform well in low dimensions, their performance seems to further increase with dimensionality. Notably, the transfer learning approach allows us to adapt the meaningful medical knowledge from a different health context.

Ultimately, since the pre-trained hyperbolic embeddings are only available in UMLS and not directly available for ICD-9 codes of our dataset, a mapping between UMLS and ICD-9 is needed. As a matter of fact, this process requires a multi-stage mapping, because the direct mapping from CUI codes in UMLS to the ICD-9 codes of the core dataset is not available. Instead, the SNOMED CT[2] were selected as an intermediary, as it serve as healthcare terminology standard and transferable both to UMLS' CUIs and ICD-9 codes.

Figure 2 provides an overview of the data flow. While the SNOMED CT were used to link the CUI with ICD-9 in order to establish a unique Poincaré embedding of each available ICD-9 code, the core dataset itself needed to be filtered for patients that have an ICD-9 record, as well. Consequently, this process reduced the original size of the dataset substantially, such that a dataset of 33k patients and 223 doctors with more than 166k interactions between them remain.

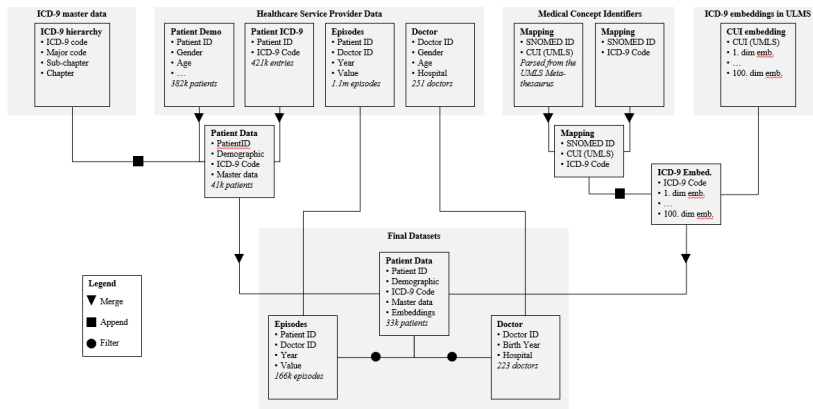

**Fig. 2.** Data diagram describing the describing the data sources, as well as the necessary mapping steps between terminologies and datasets.

## 4 Methods

### 4.1 Hyperbolic distance as a similarity measure for recommender systems

Since conventional similarity measures for recommender systems, such as cosine similarity or Pearson correlation are only inherently suited for Euclidean space [19], we introduce the notion of hyperbolic distance as a more suitable similarity measure for the data at hand. Given that each of the derived ICD-9 diseases is represented by an embedding in the hyperbolic space, the objective is to determine how similar these diseases – and ultimately the patients admitted with or doctors having treated these diseases – are with one another.

---

[2]  SNOMED CT refers to Systematized NOmenclature of MEDicine Clinical Terms that is used to encode healthcare terminology for electronic health records. All UMLS data including SNOMED CT and CUIs have been retrieved from the US National Library of Medicine (NLM).

The basic principle of hyperbolic distance as a similarity measure is simple: Once a unique embedding per either patient or doctor is derived, a patient-patient or doctor-doctor similarity score can be determined utilizing the hyperbolic distance function from equation (1). The resulting matrix of distances is subsequently scaled from 0 to 1 and subtracted from 1 in order for 0 to be the minimum similarity and 1 the maximum. Applying this heuristic yields a similarity score that is not only consistent with hyperbolic space (*i.e.,* preserving the hierarchal information and complexities of the input graph), but also as easily interpretable as conventional similarity measures. This similarity measure shall be referred to as *hyperbolic similarity* for the remainder of this paper and its implementation into the model at hand will be examined in the following section.

## 4.2 Implementation of a recommender system using hyperbolic distance

Since the ICD-9 embeddings represent metadata about patients or doctors, a RS using such embeddings can be classified as CB. While there is further data available for both patients and doctors (e.g. demographic or location data), the proposed model will consider only the ICD-9 information. In fact, we benchmark it against a conventional CB model using that very metadata for performance evaluation purposes.

As discussed in section 3, the ICD-9 information per patient from the core dataset has been enriched with the Poincaré embeddings provided by [17]. Since many patients have been admitted with more than one disease throughout their individual medical history, naturally, the majority of patients have multiple ICD-9 entries. Therefore, in order to determine a unique embedding per patient and per doctor, multiple entries need to be averaged. Due to the specific properties of the hyperbolic space, however, the usual Euclidean mean is not applicable and thus a generalisation is needed. In hyperbolic geometry, the averaging of feature vectors is done by using the Einstein midpoint [20]. The Einstein midpoint takes its simplest form in Klein coordinates and is defined as follows:

$$HypAve(x_1, \ldots, x_N) = \frac{\sum_{i=1}^{N} \gamma_i x_i}{\sum_{i=1}^{N} \gamma_i}, where \quad \gamma_i = \frac{1}{\sqrt{1-c||x_i||^2}} \ and \ c = 1 \qquad (3)$$

The Klein model is consistent with the Poincaré ball model, but since the same point has different representations in the two models, they need to be first, translated from the Poincaré to the Klein model, then averaged and ultimately mapped back into the Poincaré model in order to complete the operation. Thus, if $x_\mathbb{D}$ and $x_\mathbb{K}$ correspond to the same point in the Poincaré and the Klein model, respectively, then the following formulas serve for translating between them:

$$x_\mathbb{D} = \frac{x_\mathbb{K}}{1+\sqrt{1-c||x_\mathbb{K}||^2}} \ and \ x_\mathbb{K} = \frac{2x_\mathbb{D}}{1+c||x_\mathbb{D}||^2} \qquad (4)$$

With an appropriate methodology for hyperbolic feature vector averaging in place, a content-based model for patient-doctor matchmaking can be formulated. In formal terms, for *N* patients and *K* doctors, the patient-doctor interaction matrix $Y \in \mathbb{R}^{N \times K}$ is denoted as:

$$y_{ij} = \begin{cases} 1, & if\ patient\ i\ interacted\ with\ doctor\ j \\ 0, & otherwise \end{cases} \tag{5}$$

Adapting [15], the patient-doctor interactions are furthermore weighted with a trust measure. Thereby, the trust between a patient and a doctor is modelled by considering bot the recency and frequency of their consultation history, *i.e.,* doctors that have been visited repeatedly and recently will be weighted higher for a given patient.

Regarding feature creation, the ICD-9 embeddings need to be considered. If $V \in \mathbb{R}$ is the set of all Poincaré embeddings, with each embedding being essentially a $1 \times 100$ dimensional row vector, then for each patient *i* the set of embedding vectors is denoted as $V_i \subset V$ corresponding to all ICDs that the patient has been diagnosed with. Similarly, for each doctor *j* the set of embedding vectors is specified by $V_j \subset V$ corresponding to the ICDs of all patients that visited doctor *j*. Hence, the feature vectors of patient *i* and doctor *j* are given by the hyperbolic average of their embeddings:

$$f_i = HypAve(V_i)\ and\ f_j = HypAve(V_j) \tag{6}$$

With the feature matrices for patients and doctors established, the similarity across patients and doctors can be calculated. For purposes of simplicity, this process will be described only for doctor-doctor similarity, while it is acknowledged that the method is analogously applicable for patients. The similarity between doctor *j* and *k* is described by the above-defined *hyperbolic similarity* of their feature embeddings:

$$s_{j,k} = s_H(f_j, f_k) \tag{7}$$

Ultimately, the predicted affinity $p_{i,j}$ of a user *i* towards a doctor *j* can be computed using the following operation:

$$p_{i,j} = \frac{\sum_{k=1}^{K} y_{i,k} * s_{j,k}}{\sum_{k=1}^{K} s_{j,k}} \tag{8}$$

Recalling that *K* is equal to the total amount of doctors and $y_{i,k}$ is the trust-weighted interaction value between patient *i* and doctor *k*, it becomes evident that the predicted affinity of patient *i* is essentially given by the similarity-weighted sum of doctors the patient visited previously, divided by the sum of the weights. While RS in e-commerce usually aim to suggest primarily new, unseen items, our model does not exclude doctors the patient already interacted with for recommendation. This is of relevance insofar, as the goal of this model is to suggest the patient with the best suiting doctor for their next primary care visit, for which previously seen doctors are arguably highly relevant candidates and should by no means be excluded.

## 5 Results

While a substantial part of this paper has been dedicated to the theoretical benefits of Poincaré embeddings and their application to the given problem of patient-doctor matchmaking, it is ultimately necessary to evaluate their performance in comparison to conventional methods, in order to judge their actual value. For that purpose, we compare the following models[3]:

1. **Conventional CB**: a patient-patient-similarity based benchmark model using cosine similarity of demographic data and one-hot encoded ICD-9 data as patient features to identify patients with similar metadata,
2. **Patient ICD-9 similarity**: a patient-patient-similarity based RS using patients' averaged, hyperbolic feature vectors to identify patients with similar diseases and,
3. **Doctor ICD-9 similarity**: a doctor-doctor-similarity based RS using doctors' averaged, hyperbolic feature vectors to identify doctors that have similar expertise to the ones the patient visited in the past.

Since each of the proposed RS is presented as a sorted list – with either 3, 5 or 10 recommended doctors – it is sensible to rely on hit rate (HR) and precision (p) as evaluation criteria as the evaluation objective is to see, if the patient actually visited *one* of the recommended doctors or not. That being said, $HR@n$ refers to the number of total hits, divided by the number of patients depending on the number of recommended doctors $n \in \{3, 5, 10\}$. Analogously, $p@n$ indicates the amount of correctly predicted doctors depending on $n$. Intuitively, HR will increase with a growing number of recommendations, whereas p will decrease. As a matter of fact, the very reason to combine these two evaluation criteria is that although it is desirable to maximise the number of hits, patients should not be confused with too many options that do not meet their needs, as this might even have counterproductive effects.

Figure 3 illustrates the performance of the three suggested models regarding HR and p. While the patient ICD-similarity model is apparently not able to add substantial value scoring even slightly below the benchmark model, the doctor ICD-similarity model does, indeed, outperform the benchmark model. In fact, this allows for two major conclusions in light of the theoretical considerations in the sections above: First, hyperbolic averaging appears to be a viable method for feature averaging of Poincaré embeddings considering the substantial number of different patients and diseases doctors treat. This is insofar noteworthy, as one might reasonably assume that the more disease embeddings are being averaged, the less meaningful they become. Yet, the resulting averaged embeddings are evidently still capable of setting apart doctors fairly well. And second, the Poincaré embeddings – despite not having been trained on this dataset – can add value to this HRS. As such, these two findings show that Poincaré entity embeddings of hierarchical data are a powerful framework to help incorporate complex domain knowledge into a ML application in the healthcare sector.

---

[3] We emphasise that all proposed models are entirely CB, hence neglecting the similarity of interactions between patients or doctors. As this research is preliminary, we acknowledge that adding interaction data in a hybrid approach may boost performance substantially.

With this in mind, the business implications for the healthcare sector remain to be considered. As has been hypothesised in the introduction, the potential business value of successfully incorporating complex domain knowledge into machine learning applications in the healthcare sector may be substantial. Recalling that the goal of a match-making algorithm between patients and doctors is insofar different from typical e-commerce RS, as it aims to recommend patients with the doctor best suited for their specific, medical condition, instead of the "next best doctor", different evaluation criteria may apply from a business value perspective. For instance, one might argue that technical performance evaluation metrics such as hit rate and precision are, in fact, negligible in favour of a more qualitative evaluation. RS, in general, often suffer from popularity bias, in that they tend to suggest mostly popular doctors [21]. That being said, patients should not be matched with doctors because they are popular or because other patients with similar demographics visited them (even if this yields in high HR and p scores), but because they best fit their medical needs. Hence, we suggest for further research that recommenders akin to this work should be optimised not only with respect to hit rate and precision, since this may not fully account for popularity bias, but also towards the domain-specific quality of the recommendation.

In light of these considerations, healthcare providers can treat this factor as a value proposition for their clients. With the increasing demand for personalised healthcare solutions, RS built on patients' individual health records are arguably in-line with current market trends. Picturing a potential customer journey, the RS would suggest a patient that has been admitted with, for example, hypertension with doctors that have treated many cases of hypertension or similar diseases. In addition, making recommendations based on individual health profiles adds an explanatory perspective to the suggestions that many RS lack. Since health is a sensitive topic, in general and trust into AI solutions is a major concern in the healthcare domain, in specific, this may be a substantial driver for the success and adaptation of the recommendation engine in practice.

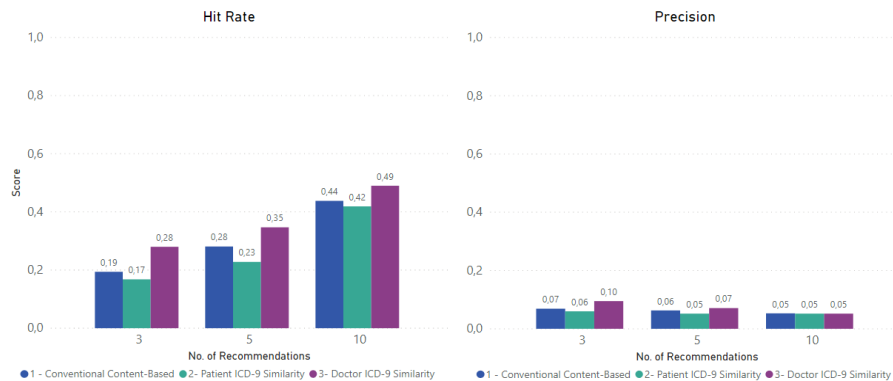

**Fig. 3.** Hit rate and precision per proposed model.

# 6    Conclusion

Overall, we demonstrate that incorporating complex domain knowledge using Poincaré embeddings of the ICD-9 hierarchy that reflect the patients' pre-existing health conditions into an HRS yields an actual performance improvement in comparison to conventional approaches . In particular, this paper examined the benefits of the hyperbolic space for representation learning tasks in theory and, furthermore, applied to real-world setting. In doing so, we show that Poincaré embeddings can contribute meaningful value in domains beyond their original scope of NLP. Moreover, we find that the incorporation of domain knowledge is of particular value in the healthcare domain, as it allows for medically personalised recommendations.

While the results of this preliminary investigation in this field are promising in principle, a set of limitations remains to be resolved in the future work. Firstly, since the proposed models are purely CB in nature, they neglect valuable information that can be retrieved from patient-doctor interaction data. Further research on a hybrid RS leveraging both interaction data and ICD-9 embeddings is a viable approach. Secondly, data consistency remains a persistent issue with a substantial portion of available data lost due to insufficient mapping between terminologies. As has been stressed before, transferability between terminologies is paramount to the further growth of AI in healthcare and healthcare analytics. Hence, the healthcare industry should continue to foster collaboration among different standardization initiatives such as SNOMED CT, UMLS, ICD, *etc*. Similar to the need of improved data consistency in the healthcare sector in general, healthcare service providers, in specific, need to drive digitization in their industry to improve the data quality as well. For instance, instead of collecting ICD information only for inpatients, all patients should be assigned with a diagnostic code in order to increase the scalability of ML solutions.

**Acknowledgements.** This work was funded by Fundação para a Ciência e a Tecnologia (UID/ECO/00124/2019, UIDB/00124/2020 and Social Sciences Data Lab, PINFRA/22209/2016), POR Lisboa and POR Norte (Social Sciences Data Lab, PINFRA/22209/2016).